\documentclass[12pt]{iopart}
\usepackage{graphicx}
\usepackage{epstopdf}
\usepackage{iopams}
\usepackage{setstack}

\begin{document}

\title[Autler-Townes effect in a dressed Jaynes-Cummings system]{Observation of Autler-Townes effect in a dispersively dressed Jaynes-Cummings system}

\author{B Suri$^{1,5}$, Z K Keane$^{1,5}$, R Ruskov$^1$, Lev S Bishop$^{2,4,5}$, Charles Tahan$^1$, S Novikov$^{1,5}$, J E Robinson$^{1}$, F C Wellstood$^{2,3,5}$ and B S Palmer$^{1,5}$}

\address{$^1$ Laboratory for Physical Sciences, College Park, MD 20740, USA}
\address{$^2$ Joint Quantum Institute, University of Maryland, College Park, MD 20742, USA}
\address{$^3$ Center for Nanophysics and Advanced Materials, University of Maryland, College Park, MD 20742, USA}
\address{$^4$ Condensed Matter Theory Center, Department of Physics, University of Maryland, College Park, MD 20742, USA}
\address{$^5$ Department of Physics, University of Maryland, College Park, MD 20742, USA}
\ead{balasuri@lps.umd.edu}

\begin{abstract}
\noindent
Photon number splitting is observed in a transmon coupled to a superconducting quasi-lumped-element resonator in the strong dispersive limit. 
A thermal population of $5.474\,$GHz photons at an effective resonator temperature of $T = 120\,$mK results in a weak $n = 1$ photon peak along with the $n = 0$ photon peak in the qubit spectrum in the absence of a coherent drive on the resonator.
 Two-tone spectroscopy using independent coupler and probe tones reveals an Autler-Townes splitting in the thermal $n = 1$ photon peak.
   The observed effect is explained accurately using the four lowest levels of the dispersively dressed qubit-resonator system and compared to results from numerical simulations of the steady-state master equation for the coupled system.
\end{abstract}

\maketitle




\section{Introduction}

Over the past decade, superconducting quantum circuits have emerged as promising candidates for quantum computation \cite{Clarke2008, Devoret2004a}. 
The coherence times of superconducting qubits have increased by several orders of magnitude through improvements in materials \cite{Chang2013}, device design and architecture \cite{Manucharyan2009, Paik2011, Barends2013}, as well as better isolation from stray noise and infrared radiation \cite{Corcoles2011, Wenner2013}. 
Many superconducting qubits are now based on cavity \cite{Paik2011,Hood2000,Raimond2001} or circuit quantum electrodynamics (cQED) \cite{Blais2004,Wallraff2004}, which rely on  the interaction of a qubit with the quantized electromagnetic field in a resonator. 
These architectures have been able to realize strong dispersive coupling between the qubit and resonator by using qubits with large dipole moments \cite{Paik2011, Koch2007}. Strong dispersive coupling \cite{Schuster2007} has, in turn, enabled the study of a variety of phenomena that were first seen in atomic systems \cite{You2011}.
 
 Here we examine photon number-splitting \cite{Schuster2007} of the spectrum of a transmon qubit \cite{ Koch2007} that is coupled to a quasi-lumped element resonator in the strong dispersive limit while the resonator is coherently driven. We also observe, in the absence of a coherent drive field on the resonator, a weak $n = 1$ photon spectral peak due to a thermal population of photons in the resonator \cite{Corcoles2011, Wenner2013, Dykman1987}. Upon pumping the system with an additional electromagnetic field at the dressed frequency of the resonator we observe a splitting of the thermal $n=1$ photon peak with a size that increases linearly as we increase the microwave drive amplitude. We show that our observations are consistent with an Autler-Townes effect \cite{Autler1955} associated with dressing of the $|\widetilde{e,0}\rangle \leftrightarrow |\widetilde{e,1}\rangle$ transition due to the strong `coupler' field. 
   While the Autler-Townes effect has previously been observed in superconducting qubits \cite{Baur2008,Sillanpaa2009}, here the effect involves the dressed resonator-qubit states and is made possible by the strong dispersive coupling. 


\begin{figure}
 \begin{center}
  {\includegraphics[width=1\columnwidth]{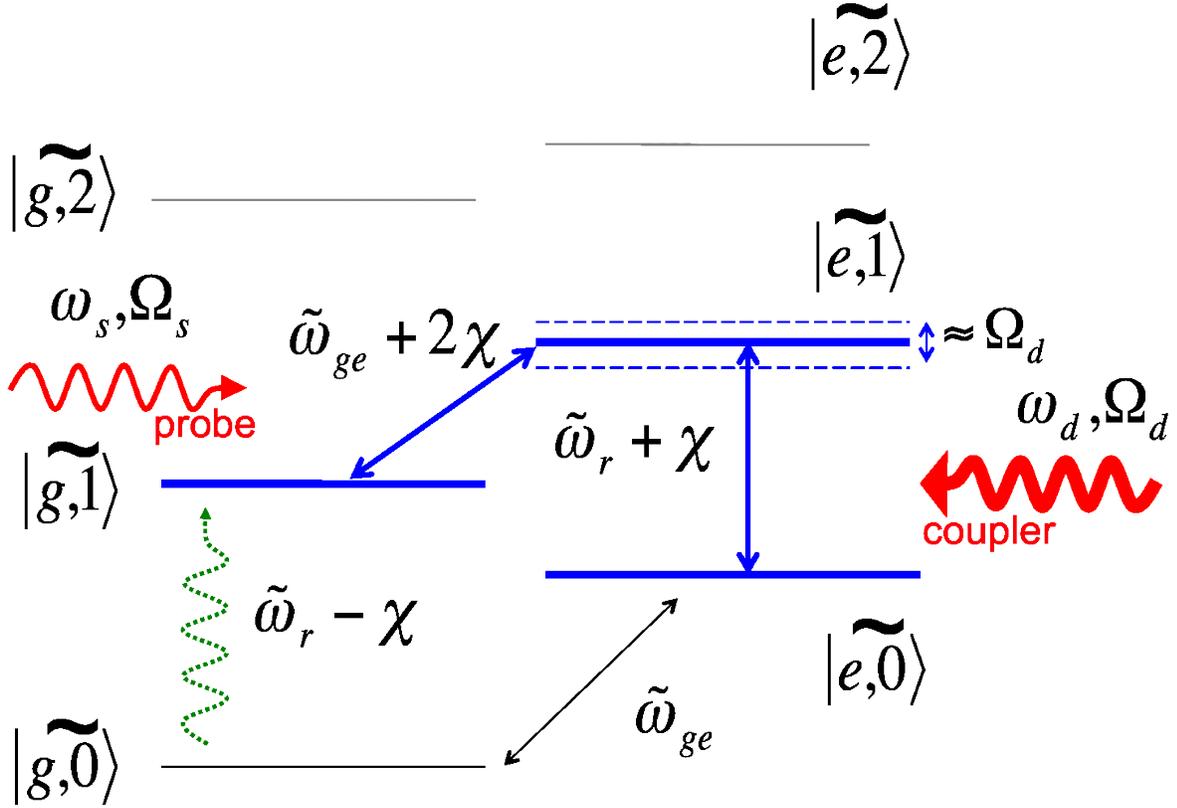}}
     \end{center}
     \caption{\textbf{Autler-Townes mechanism of the dressed qubit-resonator states}. The four lowest levels of the dispersively coupled transmon-resonator system (\ref{eqhammat}) are shown here. 
     The three levels $|\widetilde{e,0}\rangle, |\widetilde{g,1}\rangle,$ and $|\widetilde{e,1}\rangle$  form a lambda system. The $|\widetilde{e,0}\rangle \leftrightarrow |\widetilde{e,1}\rangle$ transition is strongly driven by the coupler, and the  $|\widetilde{g,1}\rangle \leftrightarrow |\widetilde{e,1}\rangle$ transition is weakly driven by the probe tone.
      The Rabi frequency $\Omega_d$ of the coupler is greater than the Rabi frequency $\Omega_s$ of the probe. 
      Therefore the dressing of the $|\widetilde{e,1}\rangle$ level is almost equal to ${\Omega_d}$. 
      In our experiment, the $|\widetilde{g,1}\rangle$ level is thermally populated (green) in the absence of a drive field.}
     \label{fig:AT}
\end{figure}

   
 \section{Theory}
 \label{sec:theory}
 \subsection{The driven Jaynes-Cummings system}
The system studied in this experiment comprises a transmon \cite{Koch2007},
which can be thought of as 
a multi-level artificial `atom'
coupled to a single harmonic mode of a superconducting  resonator.
The coupled transmon-resonator system can be modelled  
to a good approximation \cite{Bishop2010a} by a
Jaynes-Cummings Hamiltonian \cite{Jaynes1963} generalized to a multi-level atom \cite{Boissonneault2008a, Gambetta2006, Boissonneault2010}

\begin {equation}
\fl
{H_{JC} = \hbar {\omega}_r (a^{\dag} a ) + \hbar\!\!\!\!\!\! \sum_{j = \{g,e,f...\}}\!\!\!\!\!\!{\omega}_j |j\rangle \langle j| +
\hbar \!\!\!\!\!\! \sum_{j = \{g,e,f...\}}\!\!\!\!\!\!g_{j,j\!+\!1} (a^{\dag} |j\rangle \langle j\!+\!1| + a |j\!+\!1\rangle \langle j|)\,\, .}
\label{eqmlsjc}
\end{equation}
Here $\omega_r$ is the bare resonator  frequency, 
{$a^\dag\,$($a$) is the creation (annihilation) operator for the resonator mode, the} transmon states
{ $|j\rangle$ are labelled
\{${g,e,f,...}$\},
}
and $g_{j,j+1}$ is the  coupling strength
of the $|j\rangle \leftrightarrow |j+1\rangle$ transition of the transmon with the resonator mode, assuming only coupling to quasi-resonance transitions. 
This Hamiltonian can be approximately
{ diagonalized}
in the  dispersive limit \cite{Blais2004,Boissonneault2008a,Carbonaro1979},
$\Delta_{j,j+1} \equiv \omega_{j,j+1}  - \omega_r \gg g_{j,j+1}$, where $\omega_{j,j+1} \equiv \omega_{j+1}-\omega_j$
is the frequency of the $|j\rangle \leftrightarrow |j+1\rangle$ transition.
We can then write the diagonal Hamiltonian as a perturbative expansion in the
{ small parameters $\lambda_{j,j+1} \equiv g_{j,j+1}/ \Delta_{j,j+1} \ll 1$.}
In the context of our experiment, we
{ truncate}
the transmon to a two level system with ground-state $|g\rangle$ and
first-excited state $|e\rangle$. However, the finite anharmonicity of the transmon requires that we include perturbative
shifts to the energy levels of the system due to the second-excited state $|f\rangle$ \cite{Bishop2010a}.
The dispersively diagonalized Hamiltonian up to second order in $\lambda_{j,j+1}$
{ is} \cite{Blais2004, Bishop2010a}

\begin{equation}
\fl
  {\widetilde{H}^{(2)}_{JC} \approx \hbar {\widetilde{\omega}_r} (a^{\dag} a ) +
  \frac{\hbar {\widetilde{\omega}_{ge}}}{2} \sigma_z + \hbar \chi (a^\dag a) \sigma_z \, ,}
  \label{eqdisphamtls}
\end{equation}
where $\chi \simeq \chi_{ge} - \chi_{ef}/2$ is the effective dispersive shift of the resonator due to the transmon levels,
{ $\chi_{j,j+1} \equiv g^2_{j,j+1}/\Delta_{j,j+1}$ are the partial couplings,}
$\widetilde{\omega}_r \simeq \omega_r - \chi_{ef}/2$ is the dressed
resonator frequency,
$\widetilde{\omega}_{ge} \simeq \omega_{ge}+\chi_{ge}$  is  the dressed qubit transition frequency,
{ and $\sigma_z $ is the z-Pauli spin operator for the qubit.}

The last term in the Hamiltonian represents  the dispersive shift  of the resonator frequency
by an amount $\pm\chi$ depending on the qubit state.
It also  { produces an}
ac Stark shift of the qubit frequency by an amount $2\chi n$  due to
$n = \langle n|a^\dag a|n\rangle$ photons in the resonator \cite{Blais2004}.
In the strong dispersive limit the dispersive shift $\chi$ is much greater than the spectral line width
of the qubit transition ($\Gamma$) or the resonator { linewidth} ($\kappa_-$).
This enables us to resolve the Stark-shift of the qubit frequency due to { each} single photon in the resonator.
The ac Stark-shift due to each constituent Fock state $|n\rangle$ results in a separate peak in the qubit spectrum. The stationary distribution $w_n$ of the Fock states, can then be well approximated
by the relative areas under the individual photon-number peaks, with the average number of photons given by 
\begin{equation}
\fl \bar{n} = \frac{\sum_{n} \! w_n n }{\sum_{n} \! w_n } \, .
\label{eqnumphot}
\end{equation}
For a finite temperature, and in the absence of resonator driving, $w_n$ is just the usual thermal distribution \cite{Dykman1987}.
When the resonator is driven coherently from zero temperature  \cite{Glauber1963}, $w(n)$ approaches a
Poisson distribution $w^{coh}_n = e^{-\bar{n}}(\bar{n})^n / n!$.

{  We now introduce a general drive Hamiltonian with independent `coupler' and `probe' tones in the basis of dressed resonator-qubit states.}
The coupler tone { with frequency $\omega_d$} drives the `resonator-like' { dressed}  transitions
$|\widetilde{g,n}\rangle \leftrightarrow |\widetilde{g,n+1}\rangle$ and  $|\widetilde{e,n}\rangle \leftrightarrow |\widetilde{e,n+1}\rangle$.
The probe tone {with frequency $\omega_s$} drives the `qubit-like' transitions
$|\widetilde{g,n}\rangle \leftrightarrow |\widetilde{e,n}\rangle$
in the Jaynes-Cummings ladder.
The drive Hamiltonian in the  { rotating wave approximation (RWA)} is  \begin{eqnarray}
\fl H_{\rm drive} = \sum^{\infty}_{n=0} \left[ \frac{\hbar \Omega_s}{2}\bigg(P^{(n)}_{g,e} e^{i\omega_s t} +
P^{(n)}_{e,g} e^{-i\omega_s t}\bigg)  \right. \nonumber\\
\left. { }+ \sum_{k=\{g,e\}} \frac{\hbar \Omega_d}{2}\bigg(P^{(k)}_{n,n+1} e^{i\omega_d t} +  P^{(k)}_{n+1,n}e^{-i\omega_d t}\bigg)\right]
\label{eqfulldrive} \, ,
\end{eqnarray}
{\noindent  where $\Omega_s$/$\Omega_d$  are  the amplitudes of the probe/coupler,
proportional to the respective field amplitudes, ${P^{(n)}_{g,e} = |\widetilde{g,n}\rangle \langle\widetilde{e,n}|}\,$ is a  `qubit-like'   { lowering operator,} and
${P^{(k)}_{n,n+1} = |\widetilde{k,n}\rangle \langle\widetilde{k,n+1}|\,}$, with $k \in \{g,e\}$, being a `resonator-like' lowering operator.
 We note that in the strong dispersive regime, $\chi \gg \{\Gamma, \kappa_-\}$, the
`qubit-like' transitions are well resolved, and the probe drive is resonant with a unique transition. For the  `resonator-like' transitions, taking into account higher order Kerr-type nonlinearities, only a few terms survive in the summation (\ref{eqfulldrive}) and all others can be neglected in the rotating wave approximation (RWA).

\subsection{ Master equation}
\label{subsec:sim}

To simulate the system we use a density matrix formulation.
 We take into account Kerr-type nonlinearities by including
terms up to fourth order in $\lambda_{j,j+1}$.
\begin{equation}
\fl
{H^{(4)}_{Kerr}  \simeq   \hbar \zeta (a^\dag a)^2 \sigma_z + \hbar \zeta' (a^\dag a)^2} \,,
\label{eqtot4ham}
\end{equation}
where $\zeta \approx (\chi_{ef}\lambda^2_{ef} -2\chi_{ge}\lambda^2_{ge}+7\chi_{ef}\lambda^2_{ge}/4 - 5\chi_{ge}\lambda^2_{ef}/4)$
is the resonator-qubit cross-Kerr coefficient and $\zeta' \approx (\chi_{ge}-\chi_{ef})(\lambda^2_{ge}+\lambda^2_{ef})$ is
the resonator self-Kerr coefficient \cite{Gambetta2006, Boissonneault2010}.

{  The drive Hamiltonian, (\ref{eqfulldrive}), neglecting corrections of the order $\lambda^2_{j,j+1}$ in the dispersive approximation, can be written as
\begin{equation}
\fl
{H_{\rm drive} \simeq \frac{\hbar \Omega_d}{2}(a e^{i\omega_d t}+ a^\dag e^{-i\omega_d t})+
\frac{\hbar \Omega_s}{2}(\sigma^-e^{i\omega_s t}+ \sigma^+ e^{-i\omega_s t}) \, .}
\end{equation}
}
\noindent Transforming the total Hamiltonian into the rotating frame of both the drives \cite{Bishop2010a},
we obtain a time-independent  Hamiltonian,
\begin{equation}
\fl
{H_{tot} \approx \hbar \widetilde{\Delta}_d (a^\dag a)+ \frac{\hbar \widetilde{\Delta}_s}{2} \sigma_z + \hbar \chi (a^\dag a)\sigma_z +  H^{(4)}_{Kerr}+
\frac{\hbar \Omega_d}{2}(a+a^\dag) + \frac{\hbar \Omega_s}{2}(\sigma^+ + \sigma^-) \, ,}
\label{eqtotham}
\end{equation}
where $\widetilde{\Delta}_d = \widetilde{\omega}_r - \omega_d$ and
$\widetilde{\Delta}_s = \widetilde{\omega}_{ge}- \omega_s$ are the detunings of the
dressed resonator and spectroscopy tones.

Dephasing and relaxation of the qubit and losses in the resonator are accounted for in the master equation  in the Markovian approximation \cite{ Bishop2010a,Lindblad1976}.
  For the resonator, we model the dissipation of photons as a decay rate $\kappa_-$. For the transmon, we model relaxation through a
  decay rate $\Gamma_-$ and pure dephasing through the dephasing rate $\gamma_{\phi}$ in the master equation.
  In our experiment, the finite population in the qubit excited state $|e\rangle$ and the resonator $n=1$ Fock state due to finite temperature are taken into account by including excitation rates $\kappa_+$ for the resonator and  $\Gamma_+$ for the transmon,
{ where $\kappa_+/\kappa_- = \Gamma_+/\Gamma_- \simeq \exp ({-\hbar \omega_r/k_B T })$, giving us an estimate for the effective temperature of the system.}
   The master equation for the density matrix ${\rho}$ can then be written as \cite{Bishop2010a}
\begin{equation}
\fl
{\dot{\rho} = -\frac{i}{\hbar} [H_{tot} , \rho] + \kappa_- \mathcal{D}[a]\rho+\kappa_+ \mathcal{D}[a^\dag]\rho+ \Gamma_- \mathcal{D}[\sigma^-]\rho+\Gamma_+\mathcal{D}[\sigma^+]\rho} + \frac{\gamma_{\phi}}{2} \mathcal{D}[\sigma_z]\rho
\label{eqbloch}
\end{equation}
where the super-operator $\mathcal{D[{A}]{\rho}} \equiv {A\rho A^\dag - (A^\dag A \rho + \rho A^\dag A})/2$ .
We have neglected terms of the order $\lambda^2_{j,j+1}$ for each of the operators in the dissipation part of
the master equation (\ref{eqbloch}) after making the dispersive approximation.

The parameters $ \chi, \Omega_d,\Omega_s, \zeta, \zeta', \kappa_-, \kappa_+, \Gamma_-, \Gamma_+$ and $\gamma_{\phi} $
are measured experimentally (section \ref{subsec:simpar}) and  input into the simulation.
We then solve (\ref{eqbloch}) for the steady state solution $\dot{\rho} = 0$.
In our experiment, the qubit state-projective read-out signal is proportional to $\mbox{Tr}[\rho\sigma_z ]$ in the steady state.
}

\subsection{Autler-Townes mechanism}
\label{subsec:AT}

When the transition between two quantum levels is driven strongly with a resonant drive field,
{ the resulting `dressed' system can be equivalently viewed as  two split levels, with splitting equal to the Rabi frequency of the drive field (\fref{fig:AT}).
This splitting can be observed spectroscopically by probing transitions
to a third level in the system, which comprises the Autler-Townes effect  \cite{Autler1955}.
In this section, we take a closer look at (\ref{eqtotham}) to understand the mechanism of the Autler-Townes splitting of the $n=1$ photon peak observed in our experiment.} The Kerr-type nonlinearities are neglected in the context of this simple model.

{  We begin by truncating the set of basis states to the four lowest dressed levels $|\widetilde{g,0}\rangle$,  $|\widetilde{e,0}\rangle$, $|\widetilde{g,1}\rangle$ and 
$|\widetilde{e,1}\rangle$ (\fref{fig:AT}). In the experiment, the $|\widetilde{e,0}\rangle \leftrightarrow |\widetilde{e,1}\rangle$
transition is driven by a strong `coupler' field with
strength $\Omega_d$ and detuning $(\widetilde{\omega}_r +\chi) - \omega_d \equiv \widetilde{\Delta}_d+\chi $,
and the $|\widetilde{g,1}\rangle \leftrightarrow |\widetilde{e,1}\rangle$
is `probed' by a weak field of strength  $\Omega_s$  and  detuning  $(\widetilde{\omega}_{ge} + 2\chi) - \omega_s \equiv \widetilde{\Delta}_s +2\chi$
(figure \ref{fig:AT}).

In this truncated basis, the Hamiltonian can be represented by the matrix
\begin{equation}
\fl
{H^{(2)}_{\rm 4-levels} \approx
\hbar  \left(
 \begin{array}{cccc}
 -\widetilde{\Delta}_s/2 & 0&0&0\\
 0&\widetilde{\Delta}_s/2&0&\Omega_d/2\\
 0&0&\widetilde{\Delta}_d-\widetilde{\Delta}_s/2-\chi&\Omega_s/2\\
 0&\Omega_d/2&\Omega_s/2& \widetilde{\Delta}_d + \widetilde{\Delta}_s/2+\chi\\
  \end{array}
  \right) }\, ,
  \label{eqhammat}
\end{equation}
\noindent where we have assumed the RWA and excluded transitions detuned from the drives.
When the probe and coupler drives are exactly resonant with the respective transitions, the detunings  obey $\tilde{\Delta}_s+2\chi = 0$ $\tilde{\Delta}_d+\chi = 0$.
Under this condition the  eigenvalues of the Hamiltonian in (\ref{eqhammat}) 
show a splitting equal to 
\begin{equation} 
\fl \delta = (\Omega^2_d + \Omega^2_s)^{1/2} \, ,
\label{eqsplitting}
\end{equation}
and for
$\Omega_s \ll \Omega_d$, this gives an Autler-Townes splitting linear in the coupler Rabi frequency
$\delta \simeq \Omega_d \sim (P_{rf})^{1/2}\,$, where $P_{rf}$ is the drive power of the coupler tone.
This simple model for the splitting offers good quantitative agreement with the experiment (section \ref{subsec:ATdata}).


\section{Experiment}
\label{sec:exp}
\subsection{Device design and fabrication}
\Fref{fig:device} shows the transmon \cite{Koch2007} coupled to a superconducting lumped-element resonator \cite{Kim2011}, which is in turn coupled to a coplanar waveguide transmission line used for excitation and measurement. The transmon is formed from two Josephson junctions  (junction area $\approx 100 \times 100\ $ nm$^2$) shunted by an interdigitated capacitor with finger widths of 10 $\mu$m, lengths of 70 $\mu$m, and separation between fingers of 10 $\mu$m. The junctions are connected to form a superconducting loop with  a nominal loop area of  $4 \times 4.5\, \mu$m$^2$. The Josephson junction loop is placed close to a shorted current bias line to finely tune the critical current of the parallel junctions and hence the transition frequency of the qubit. 

The device was fabricated by depositing a $100 \,$nm thin film of aluminium by thermal evaporation on a commercial \textit{c}-plane sapphire substrate.  The resonator, transmission line, and on-chip flux bias line were patterned using photolithography and etched with a standard aluminium etchant. The transmon was subsequently patterned using electron-beam lithography and the Al/AlO$_{x}$/Al tunnel junctions were formed by  double-angle evaporation with an intermediate oxygen exposure step.

           
\begin{figure}
 \begin{center}
      \includegraphics[width=1\columnwidth]{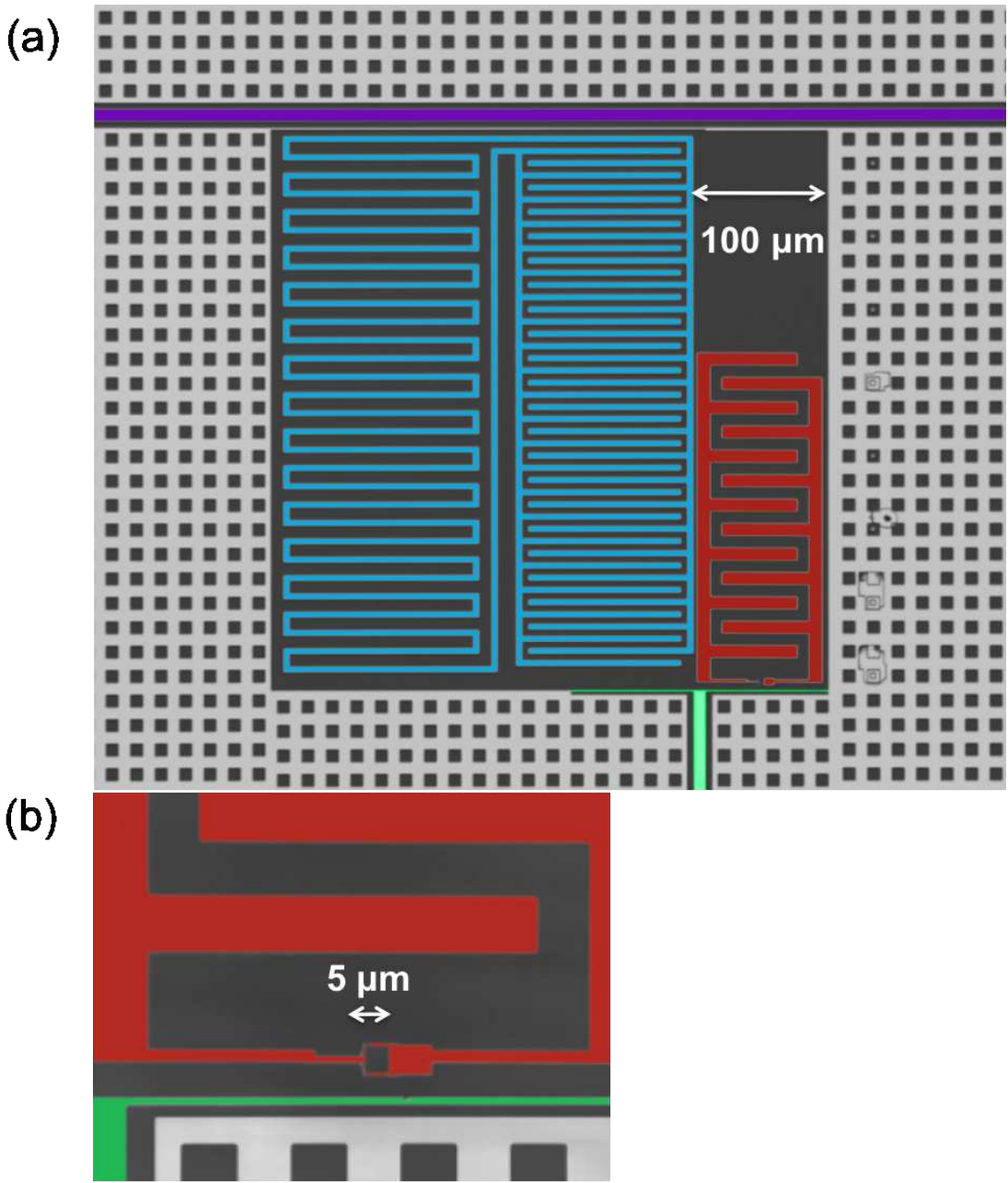}
     \end{center}
     \caption{\textbf{Colourized micrograph of device}.
       \textbf{(a)} A lumped element resonator (blue) and transmon (red) are coupled to a coplanar waveguide transmission line (violet) and surrounded by a perforated ground plane (white). The resonator consists of a meandering inductor and an interdigitated capacitor. The transmon has two Josephson junctions in parallel the allow the transition frequency to be tuned with an external magnetic field and an on-chip flux bias (green). \textbf{(b)} Detailed view of transmon's Josephson junctions and flux bias line. }
            \label{fig:device}
\end{figure}

 
\subsection{Experimental set-up}
The device was mounted in a  hermetically sealed copper box and attached to the mixing chamber of an Oxford Kelvinox 100 dilution refrigerator with a base temperature of $20\,$ mK. To isolate the device from thermal noise at higher temperatures, the input microwave line to the device has a 10 dB attenuator mounted at 4 K, 20 dB at 0.7 K, and 30 dB  at 20 mK on the mixing chamber. On the  output (\textit{i.e}. read-out) microwave line, two 18 dB isolators with bandwidths from 4 to 8 GHz are placed in series at 20 mK. The output microwave signal then goes through a 3 dB attenuator and a high electron mobility transistor amplifier at 4 K before being amplified further at room temperature and mixed down to an intermediate frequency (IF) of 10 MHz. 

Three microwave sources are used in the measurement: probe, coupler, and read-out. The read-out and probe tones are pulsed on and off, while the coupler tone is applied continuously for the duration of the measurement. To measure the excited-state population of the transmon, a high power (\textit{i.e}. 50 dB larger than the power of the coupler tone) at a frequency corresponding to the bare cavity is applied \cite{Reed2010}. This read-out relies on the Jaynes-Cummings nonlinearity of the coupled resonator-qubit system \cite{Boissonneault2010,Bishop2010} and depending on the state of the system, either a large transmissivity is observed (\textit{e.g}. $|g\rangle$ state) or  small transmissivity (\textit{e.g}. $|e\rangle$ state). 
 
For spectroscopic measurements, a probe pulse is first applied for a duration of $5\, \mu$s at an amplitude just large enough to saturate the qubit transition without causing large power broadening. A read-out pulse of duration $5 \,\mu$s is applied $20\,$ns after turning off the probe. The mixed down IF signal is digitized using a data acquisition card and the in-phase and quadrature components are demodulated before being recorded. 
 
\subsection{Experimental parameters}
\label{subsec:simpar}
	From spectroscopic and time-domain measurements the main system parameters were determined. The resonator has a bare resonant frequency $\omega_r/2\pi = 5.464\,$GHz, internal quality factor $Q_I = 190,000$, and a loaded quality factor $Q_L \equiv \omega_{r}/\kappa_{-}= 18,000$. The  parallel resistance of the Josephson junctions yielded a  maximum Josephson energy $E_{J,max}/h \approx 25\,$GHz and the transmon has a Coulomb charging energy  of  $E_{c}/h = 250\,$MHz. This gives a maximum ground-to-first excited state transition frequency, $\omega_{ge,max}/2\pi \simeq (\sqrt{ 8E_{J,max} E_c} - E_c)/h = 7.1\,$GHz for the qubit. The qubit transition frequency was tuned using a combination of an external superconducting magnet and  the on-chip flux bias to $\widetilde{\omega}_{ge}/2\pi = 4.982\,$GHz, which corresponds to a detuning of $\Delta_{ge} /2\pi= 482\,$ MHz from the resonator. From spectroscopic measurements, we determined the effective dispersive shift $\chi/2\pi = 4.65\,$MHz, and the parameters $\chi_{ge}/2\pi = -10\,$ MHz and $\chi_{ef}/2\pi = -10.7\,$MHz. From the definitions of $\chi_{ge}$ and $\chi_{ef}$, we determine $g_{ge}/2\pi = 70\,$MHz and $g_{ef}/2\pi = 89\,$ MHz. The resonator self-Kerr coefficient $\zeta'/2\pi = 85\,$kHz and the transmon-resonator cross-Kerr coefficient $\zeta/2\pi = 23\,$kHz are then determined from (\ref{eqtot4ham}). 

Time-domain coherence measurements performed at this resonator-qubit ($\Delta_{ge} /2\pi= 482\,$ MHz) detuning revealed a qubit relaxation time $T_{1} = 1/(\Gamma_{-} +\Gamma_{+})  = 1.6 \, \mu$s$^{-1}$ and a Rabi decay time  $T' = 1.6\,\mu$s for the first excited state of the qubit. From these measurements, the pure dephasing rate is estimated at $\gamma_{\phi}  \equiv 1/T_\phi \approx 2\times10^5\,$s$^{-1}$. 
    
\section{Results}
\label{sec:datasim}
\subsection{Photon number-splitting}
Figure \ref{fig:Numsplitsameaxis}(a) shows  spectroscopy of the transmon with no drive field applied to the resonator. The spectrum shows the dressed qubit ground-to-first excited state transition at $\widetilde{\omega}_{ge}/2\pi = 4.982\,$GHz. The smaller spectroscopic peak detuned  by -9.3 MHz at  $(\widetilde{\omega}_{ge}-2\chi)/2\pi =4.973\,$GHz is due to  one  $\widetilde{\omega}_{r} -\chi/2\pi=5.474\ $GHz photon occasionally being present in the resonator \cite{Dykman1987} from thermal excitations. From the relative areas under the two spectral peaks, we estimate a fractional thermal population of $n_{th} = 0.1$ photons, corresponding to a temperature of about $120\,$mK for the resonator. This effective temperature is much higher than the base temperature $\sim 20\,$mK of the dilution refrigerator, possibly due to a leakage of higher frequency photons \cite{Corcoles2011, Wenner2013}. 

\begin{figure}
 \begin{center}
      \includegraphics[width=1\columnwidth]{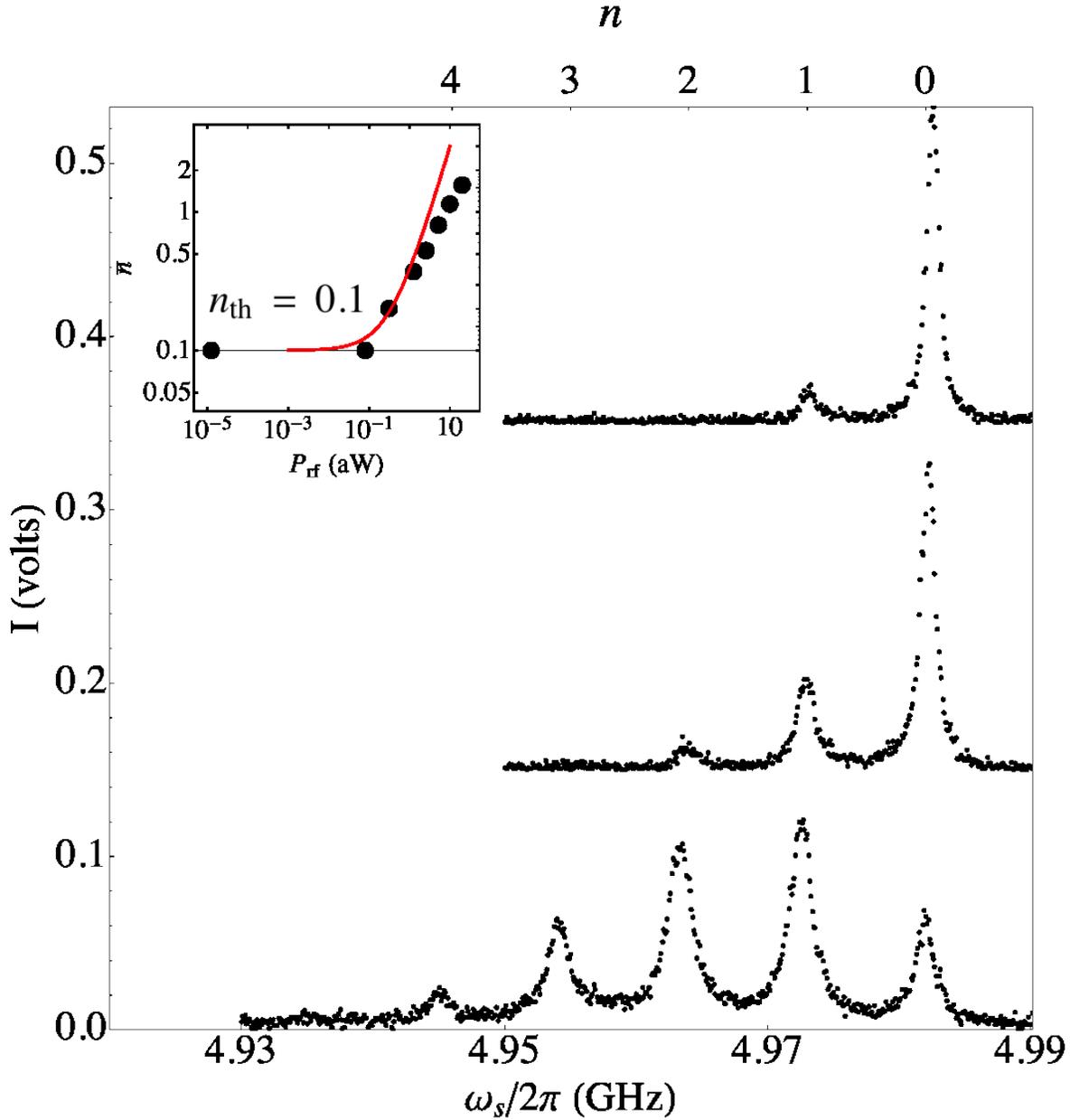}
     \end{center}
     \caption{ \textbf{Photon number-splitting in Transmon Spectrum.} In-phase signal versus probe frequency ($\widetilde{\omega}_{ge}/2\pi$) with a drive amplitude of $\Omega_{s}/2\pi = 300\ $kHz. \textbf{(a)}  Spectroscopy with no coupler tone applied. The qubit ground-to-first excited state transition frequency is seen at $\widetilde{\omega}_{ge}/2\pi = 4.982\,$GHz. The spectroscopic peak detuned by  $2\chi = -2\pi(9.3)\,$MHz at $4.973\,$GHz is due to a thermal population $n_{th} =  0.1$ photons in the resonator. \textbf{(b)} A coupler tone applied at $\tilde{\omega}_d/2\pi = 5.474\,$GHz and power  $P_{rf} = 1.25\,$aW produces a population with an average of  $\bar{n} = 0.35$ photons stored in the resonator. \textbf{(c)} A coupler tone applied at $\tilde{\omega}_d/2\pi = 5.474\,$GHz and power  $P_{rf} = 25\,$aW produces a population with an average of  $\bar{n} = 1.5$ photons stored in the resonator. \textbf{Inset}: Average number of photons  using (\ref{eqnumphot}) stored in the resonator versus coupler power in the weak driving limit. At very low drive powers  ($P_{rf} < 0.1\,$ aW), the average photon number $\bar{n}$ plateaus to $n_{th} = 0.1$. The red curve is a model  consisting of a thermal population plus a contribution due to a coherent population. Deviations from this linear relation can be seen as the applied power increases.}
     \label{fig:Numsplitsameaxis}
\end{figure}

Upon driving the resonator with a  coupler tone at $\omega_d/2\pi = 5.474\,$GHz,  which is resonant with the transition $|\widetilde{g,0}\rangle \leftrightarrow |\widetilde{g,1}\rangle$ (\fref{fig:AT}), we increase the mean occupancy of the resonator from its equilibrium value and create a coherent state.
 Since the coherent state is a superposition of Fock states, the qubit spectrum has multiple peaks, one for each Fock state.
 When applying a power of $P_{rf} =  1.25\,$aW at the dressed resonator frequency (figure \ref{fig:Numsplitsameaxis}(b)), an increase in the height of the $\widetilde{\omega}_{ge}-2\chi$ peak is observed and a spectral peak at $\widetilde{\omega}_{ge}-4\chi$ begins to appear. The peak at $\widetilde{\omega}_{ge}$ is still the largest. \Fref{fig:Numsplitsameaxis}(c) shows the spectrum for an applied resonator drive power of 25 aW. In this case, more spectral peaks are observed and the $\widetilde{\omega}_{ge}-2\chi$ is the largest.

We can calculate the average number of photons ($\bar{n}$) by fitting the spectral peak associated with each Fock state and using the relative peak areas to weight each Fock state (\ref{eqnumphot}). The relative weights are also found to follow a Poisson distribution once the resonator is pumped into a coherent state \cite{Schuster2007}.  The inset of Figure~\ref{fig:Numsplitsameaxis} shows the average number of photons versus the applied power in the weak driving limit. For very weak driving $P_{rf} < 0.1\,$aW, the thermal photon population $n_{th} = 0.1$ is the dominant contribution to $\bar{n}$. Above an applied power of  $P_{rf} > 0.1\,$aW $\bar{n}$ monotonically increases. For small applied powers in this region, $\bar{n} = (2 Q_{L}/Q_{C})P_{rf}/(\hbar \tilde{\omega}_r\kappa_{-})$  where the first term renormalizes the power applied to the transmission line to the power stored in the resonator and $Q_{C}$ is the quality factor due to external coupling. Using this linear relation and the excess photon number population from the applied coupler tone in figure  \ref{fig:Numsplitsameaxis}(b) an attenuation of  $\alpha = 65\,$dB is calculated for the input microwave line. The red curve in the inset is a model for $\bar{n}$ consisting of a contribution from a power independent thermal contribution plus a coherent state population with an applied linear power dependence. As can be seen in this figure, we see a strong deviation from linear behavior.  It is important to note that the average occupancy of the resonator $\bar{n}$ is, in general, expected to be a nonlinear function of $P_{rf}$, given by
 \begin{equation}
 \fl
 \bar{n}_{\pm} = \frac{Q_{C}}{2Q_{L}} \frac{\kappa_- P_{rf}}{\hbar \tilde{\omega}_r}\frac{1}{\left((\kappa_-/2)^2 + (\tilde{\Delta}_d \pm \chi )^2\right)} \, ,
 \end{equation}
 \noindent where the $\pm$ stands for the qubit being in the ground or excited state respectively $ \cite{Gambetta2006}$. A detailed analysis of the full nonlinear relation is beyond the scope of this paper.
  
Figure~\ref{fig:numsplitvsdetuning} shows  two-tone spectroscopy of the qubit as we vary the frequency of the coupler tone. 
In this plot, the vertical bands at frequencies  $\omega_{s}/2\pi = 4.982\,$GHz and 4.973 GHz are just the $n=0$ and $n=1$ photon peaks as shown in figure~\ref{fig:Numsplitsameaxis}.
The prominent diagonal band seen between the $n=0$ and $n=1$ photon bands in this figure corresponds to a two-photon `blue' sideband transition from the $|\widetilde{g,0}\rangle$  to $|\widetilde{e,1}\rangle$ \cite{Wallraff2007};  the sum of the frequencies ($\omega_s,\omega_d$) of the drive photons along this diagonal band is equal to the corresponding transition frequency. In general, a diagonal band with slope $-1/n$ should appear when the detunings of the drives satisfy  $\widetilde{\Delta}_s-n\widetilde{\Delta}_d = n\chi$ corresponding to the  sideband transition  $|g,n-1\rangle \leftrightarrow |e,n\rangle$.
 For example, the sideband transition  $|\widetilde{g,0}\rangle \leftrightarrow |\widetilde{e,1}\rangle$  level appears as a band of slope $-1$ and the $|\widetilde{g,1}\rangle \leftrightarrow |\widetilde{e,2}\rangle$ transition appears as a faintly visible band of slope $-1/2$ in figure \ref{fig:numsplitvsdetuning}.

\begin{figure}
\begin{center}
\includegraphics{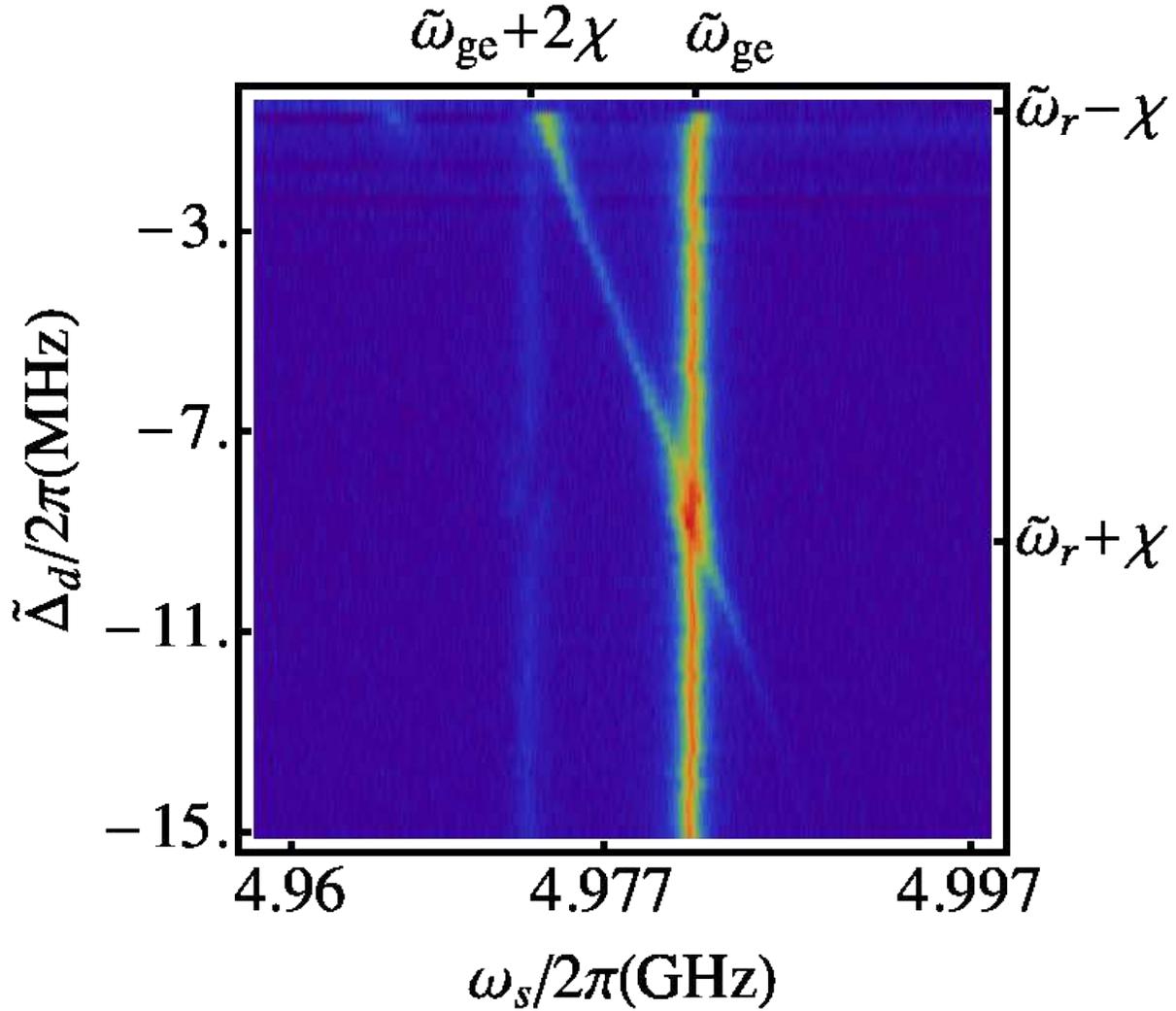}
     \end{center}
 \caption{ \textbf{Two-tone spectroscopy of the transmon-resonator system.} Colour plot of qubit excited state versus probe ($\omega_{s}$) frequency and detuning of the coupler drive at a coupler power of $38\,$aW. Red corresponds to increase in the population of the excited state of the qubit. The qubit transition is seen as a vertical band at $\omega_s/2\pi = \widetilde{\omega}_{ge}/2\pi=  4.982\,$GHz. The spectral peak due to thermal photon excitations is seen as a vertical band at $4.973\,$GHz. The diagonal band between $n=0$ and $n=1$ photon peaks is due to the sideband two-photon transition from $|\widetilde{g,0}>$ to $|\widetilde{e,1}\rangle$. An analogous sideband transition $|\widetilde{g,1}\rangle \leftrightarrow |\widetilde{e,2}\rangle$ is faintly visible with of slope -1/2 towards the top-left portion of the figure. Note also the small Autler-Townes splitting of the thermal-photon peak at  $(\widetilde{\omega}_{ge} +2\chi, \widetilde{\omega}_r +\chi)$ in the plot. }
     \label{fig:numsplitvsdetuning}
\end{figure} 

\subsection{Autler-Townes splitting in the dressed Jaynes-Cummings system}
\label{subsec:ATdata}
Another feature of the data in figure \ref{fig:numsplitvsdetuning} is a small splitting in the thermal $n=1$ photon peak when the frequencies of the probe and coupler tones are ($\omega_s,\omega_d) \simeq (\widetilde{\omega}_{ge}+2\chi,\widetilde{\omega}_r+\chi$). We examine this splitting more closely in this section.
Here we follow the convention of section \ref{subsec:AT} and refer to the strengths of the probe and coupler tones in terms their Rabi frequencies $\Omega_s$ and $\Omega_d$. The Rabi frequency of the coupler tone $\Omega_d$ can be independently calibrated from the power of the coupler tone using the relation
\begin{equation}
\fl
\Omega_d \simeq \left( \frac{Q_C\kappa_- P_{rf}}{2 Q_L\hbar \tilde{\omega}_r} \right)^{1/2} \, ,
\label{eqrabicalib}
\end{equation}
while the Rabi frequency $\Omega_s$ of the probe tone was measured independently from the Rabi oscillations of the qubit.  

\Fref{fig:DatSim}(a)-(c) shows measurements of the splitting in the thermal $n = 1$ photon peak as we vary the strength of the coupler tone  while keeping the strength of the probe tone fixed  $\Omega_s/2\pi \simeq 0.3\,$MHz. \Fref{fig:DatSim}(a) shows that when we apply a coupler tone with strength $\Omega_d/2\pi \simeq 1.3\,$MHz, we observe a splitting in the thermal $n=1$ photon peak with a splitting size that is almost equal to $\Omega_d$. In figure~\ref{fig:DatSim}(b), we lower the strength of the coupler to $\Omega_d/2\pi \simeq 1\,$MHz and we notice a corresponding decrease in the splitting size to $\sim 1\,$MHz. Upon further lowering the strength of the coupler to $\Omega_d/2\pi \simeq 0.6\,$MHz, the splitting decreases in size to $\sim 0.6\,$MHz. 

The splitting size can be understood based on an Autler-Townes mechanism (section \ref{subsec:AT}) shown in figure \ref{fig:AT}. To begin with, the $|\widetilde{g,1}\rangle$ level is populated due to thermal excitation of photons in the resonator \cite{Corcoles2011, Wenner2013}. Subsequently the probe and coupler tones are applied on resonance with the transitions $|\widetilde{g,1}\rangle \leftrightarrow |\widetilde{e,1}\rangle$ and $|\widetilde{e,0}\rangle \leftrightarrow |\widetilde{e,1}\rangle $ (see section \ref{subsec:AT}) respectively. In the presence of the two drive fields, the $|\widetilde{e,1}\rangle$ level splits into a pair of levels separated by $\delta = (\Omega^2_s + \Omega^2_d)^{1/2}$ (\ref{eqsplitting}). 
    In the limit $\Omega_d \gg \Omega_s$ this splitting is almost equal to $\Omega_d$ and is observed spectroscopically upon probing the  $|\widetilde{g,1}\rangle \leftrightarrow |\widetilde{e,1}\rangle$ transition.
 
The simple model in (\ref{eqhammat}) also predicts the overall `shape' of the splitting in the $(\omega_s,\omega_d)$ plane in \fref{fig:DatSim}(a)-(c). When $\tilde{\Delta}_s + 2\chi =0$ and $\tilde{\Delta}_d + \chi \neq 0$, only the 
$|\widetilde{g,1}\rangle \leftrightarrow |\widetilde{e,1}\rangle$ transition is resonantly driven, while the influence of level $|\widetilde{e,0}\rangle$ is energy suppressed.
This explains the vertical band corresponding to the $n=1$ photon peak at $4.973\,$GHz.
When $\tilde{\Delta}_s+2\chi = \tilde{\Delta}_d +\chi \neq 0$, the difference of the drive frequencies corresponds to the 
$|\widetilde{e,0}\rangle \leftrightarrow |\widetilde{g,1}\rangle$ transition. This two-photon `red' sideband transition \cite{Wallraff2007} explains the slope $+1$ of the splitting in the \fref{fig:DatSim}.

\begin{figure}
 \begin{center}
	       \includegraphics[width=1\columnwidth]{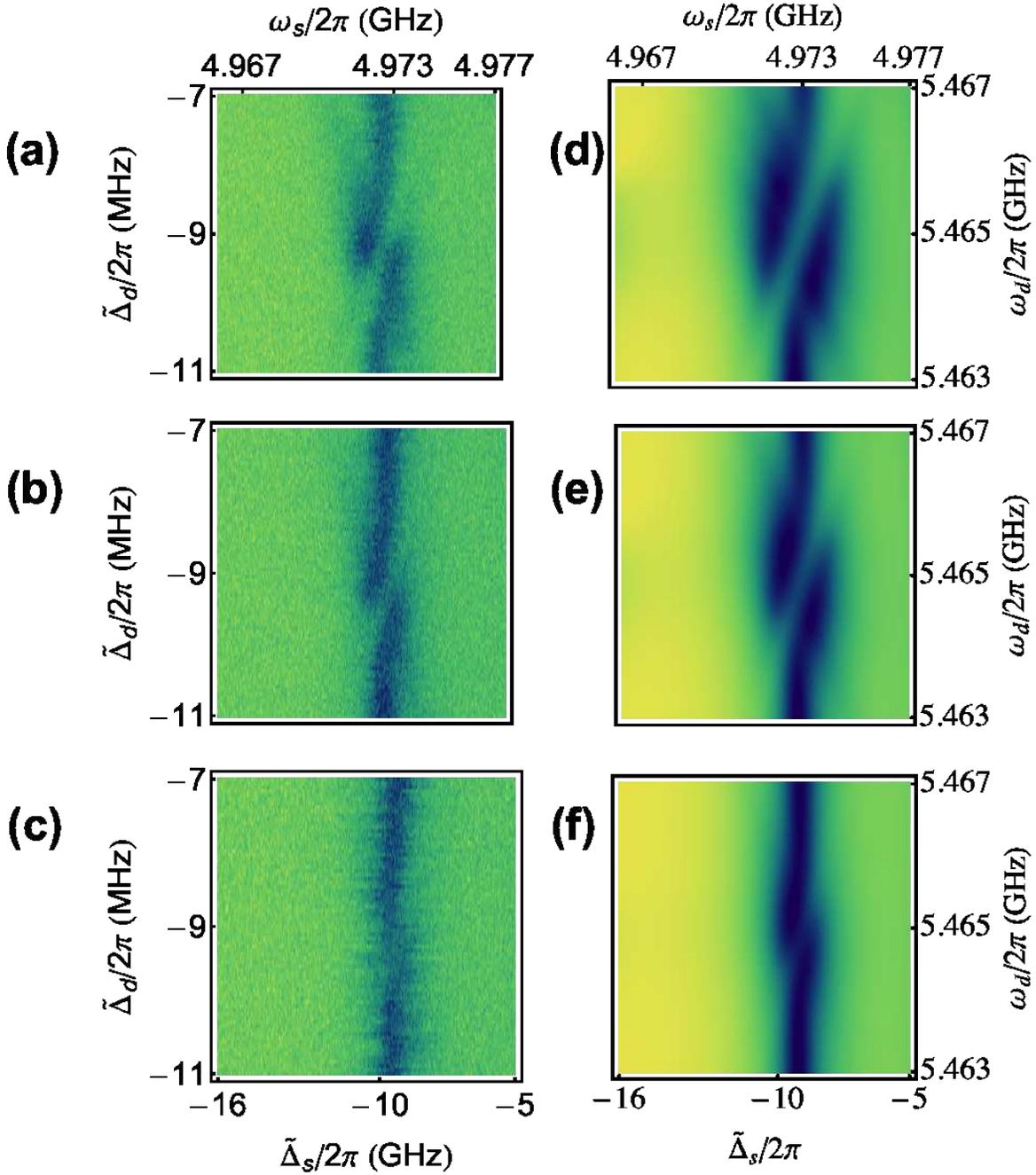}
     \end{center}
      \caption{ \textbf{Two-tone spectroscopy of Autler-Townes splitting}. Colour plots of measurements of the qubit excited state versus frequency of the applied probe ($\omega_{s}$) and coupler ($\omega_{d}$) on the $n=$ thermal photon peak.
     \textbf{(a)-(c)}  Data showing the Autler-Townes splitting of the thermal $n=1$ photon peak for three different strengths of the coupler drive.  Splitting for a coupler amplitude of \textbf{(a)} $\Omega_d/2\pi \sim 1.3\,$MHz, \textbf{(b)} $\Omega_d/2\pi \sim 1\,$MHz, and \textbf{(c)}  $\Omega_d \sim 0.6\,$MHz.  \textbf{(d)-(f)} Results from numerical simulations of the steady-state master equation for the resonator-transmon system. Plot of $\mathrm{Tr(\rho \sigma_z)}$ for Autler-Townes splitting in the thermal $n=1$ photon peak for amplitudes of the coupler drive \textbf{(d)} $\Omega_d/2\pi = 1.3\,$MHz, \textbf{(e)} 1 MHz, and \textbf{(f)} 0.6\ MHz. 
     }
     \label{fig:DatSim}
\end{figure}

Figure \ref{fig:DatSim}(d)-(f) shows simulations of the system-bath master equation at the steady state (section \ref{subsec:sim}) in the region around the thermal $n=1$ photon peak when driving the coupler with a strength of $\Omega_d/2\pi$ = 1.3 MHz (d), 1 MHz (d), and 0.6 MHz (f). 
The parameters $\chi, \Omega_d,\Omega_s, \zeta, \zeta', \kappa_-, \kappa_+, \Gamma_-, \Gamma_+$ and $ \gamma_{\phi} $ in the master equation were determined from independent experiments (section \ref{subsec:simpar}) and equation (\ref{eqbloch}) was then solved for $\rho$ for the steady state solution. 
 Here we plot $\mathrm{Tr (\rho. \sigma_z)}$ to simulate the qubit state-projective read-out in our experiment \cite{Boissonneault2010,Reed2010, Bishop2010}.
We included the two lowest qubit levels $|g\rangle, |e\rangle$, the ten lowest resonator levels, the resonator self-Kerr and qubit-resonator cross-Kerr terms in the simulation. The population in the $n=1$ Fock state of the resonator and the $|e\rangle$ state of the transmon due to finite temperature were taken into account through the excitation rates $\kappa_+ \approx \kappa_-/10$, $\Gamma_+ \approx \Gamma_-/10$  in the system master equation (\ref{eqbloch}). 

 \begin{figure}
 \begin{center}
      \includegraphics[width=1\columnwidth]{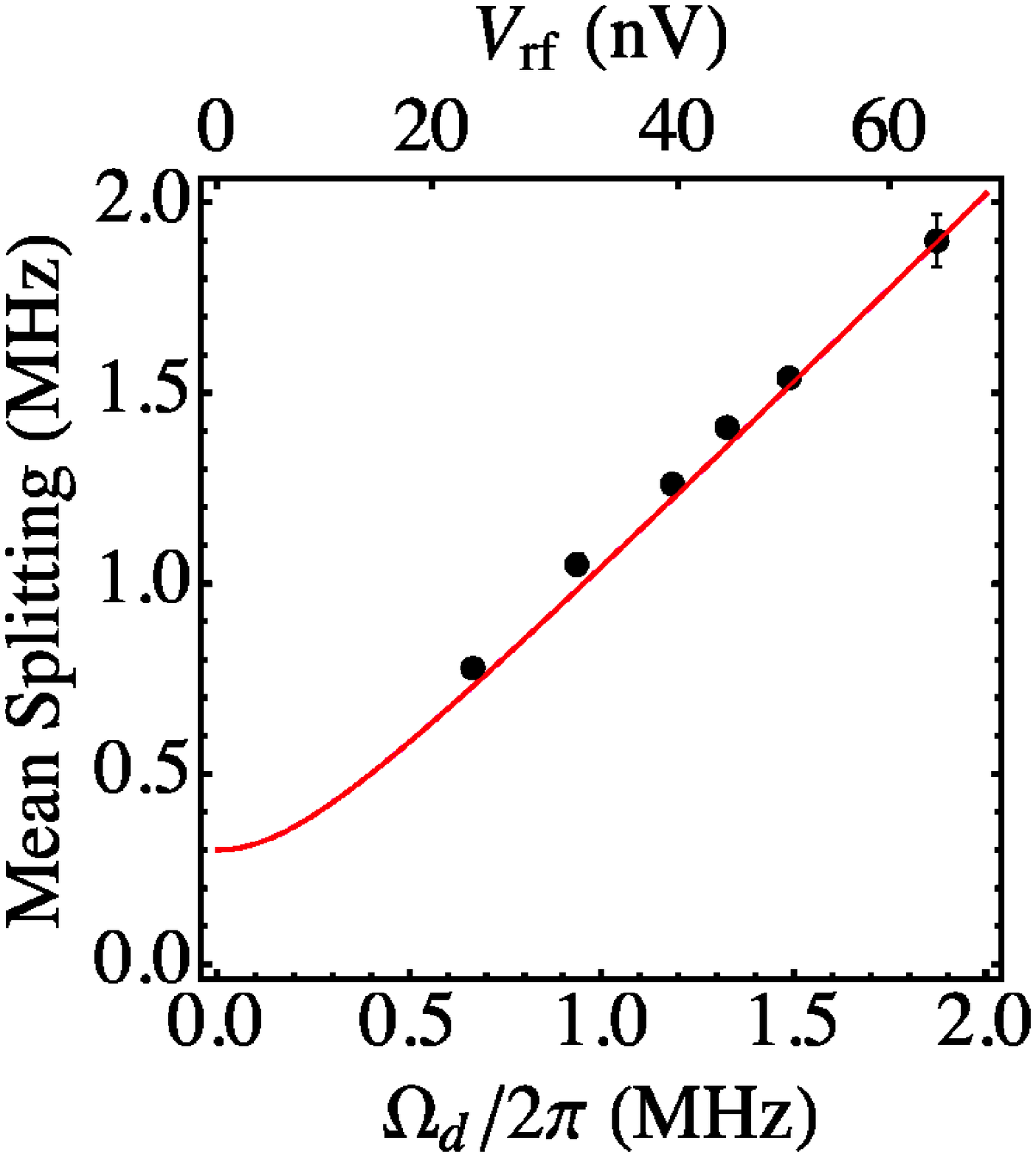}
     \end{center}
     \caption{ \textbf{Autler-Townes splitting size versus drive amplitude of the coupler field.} Measured Autler-Townes splitting (black circles) in the thermal $n=1$ photon peak increases linearly with the voltage of the coupler drive $V_{rf} = (P_{rf})^{1/2}$. The Rabi frequency $\Omega_d$ of the coupler is calibrated from $P_{rf}$ from (\ref{eqrabicalib}). The red curve is a model of the splitting size (\ref{eqsplitting}) where $\Omega_s/2\pi = 0.3\,$MHz was calibrated from Rabi oscillations and $\Omega_d$ was calibrated from data in \fref{fig:numsplitvsdetuning}. }
     \label{fig:SplittingvsRabi}
\end{figure}

Finally, figure \ref{fig:SplittingvsRabi} shows the experimentally measured splitting vs  the Rabi frequency of the coupler. The Rabi frequency of the coupler $\Omega_d$ was calibrated independently from the power of the coupler drive using (\ref{eqrabicalib}). The root-mean-square voltage $V_{rf} \propto (P_{rf})^{1/2}$ of the coupler at the device is calculated from $P_{rf}$  assuming a 50$\, \Omega$ impedance for the transmission line. The red curve in the \fref{fig:SplittingvsRabi} is the splitting size $\delta/2\pi \simeq(\Omega^2_d+\Omega^2_s)^{1/2}/2\pi$ predicted by the simple model (\ref{eqsplitting}) for a fixed $\Omega_s/2\pi \simeq 0.3\,$MHz and it agrees well with the experimental data. Here we see that the observed Autler-Townes splitting is nearly linear in the amplitude of the coupler drive voltage $V_{rf}$ and is almost equal to the Rabi frequency of the coupler field, as expected from (\ref{eqsplitting}). 

  \section{Conclusion}
 In conclusion, we studied photon number-splitting in the spectrum of a  transmon coupled to a lumped element resonator in the strong dispersive regime. 
 In the absence of a coherent field driving the resonator, we observed an average thermal population $n_{th} = 0.1$ of microwave photons in the resonator, corresponding to an effective temperature of $120\,$mK.
 We observed additional photon number splitting of the qubit spectrum when a coherent coupler tone is applied to the resonator. 
 In the presence of a strong coupler field and a weak probe field, we observed an Autler-Townes spitting of the thermal $n=1$ photon peak. 
The size of the splitting increased linearly with the amplitude of the coupler tone as expected for an Autler-Townes effect in the  `lambda' system comprising the $|\widetilde{g,1}\rangle,\, |\widetilde{e,0}\rangle,\, |\widetilde{e,1}\rangle $  levels of the dressed Jaynes-Cummings transmon-resonator system. 
 Numerical simulations of the steady state  system-bath master equation for the density matrix with two qubit levels and up to ten resonator levels agree well with the experimental observations. 
  
  We note that our simple model also predicts an Autler-Townes splitting of the $n=0$ photon qubit peak when the probe and coupler frequencies are $(\omega_s, \omega_d) \simeq (\tilde{\omega}_{ge}, \tilde{\omega}_{r}+\chi)$. 
  We see a hint of this splitting in figure~\ref{fig:numsplitvsdetuning} but it is not fully resolved because of the line-width of the qubit $|\widetilde{g,0}\rangle \leftrightarrow |\widetilde{e,0}\rangle$ transition. 
The Autler-Townes mechanism for the splitting in the $n=0$ photon peak involves the $|\widetilde{g,0}\rangle,\, |\widetilde{e,0}\rangle,\, |\widetilde{e,1}\rangle $ levels in figure~\ref{fig:AT}. The three-level subsystem is operated in a `ladder' configuration in this case, with the $|\widetilde{e,0}\rangle \leftrightarrow |\widetilde{e,1}\rangle$ transition driven strongly with a `coupler' tone and the $|\widetilde{g,0}\rangle \leftrightarrow |\widetilde{e,0}\rangle$ transition weakly `probed'.

In atomic systems, observing the Autler-Townes effect is often a precursor to seeing electromagnetically induced transparency or absorption (EIT/EIA) \cite{Boller1991, Ian2010}. 
 EIT (EIA) gives rise to exotic phenomena such as `slow-light' \cite{Hau1999}, and its use as a sensitive probe for the decoherence of a quantum state was proposed \cite{Murali2004}.
 Clear observation of EIT in superconducting systems typically requires engineering the coherence of the three `atom' states independently \cite{Murali2004}, and how this can be done is an open question in superconducting qubit research.
Ian \textit{et al.} \cite{Ian2010} have theoretically explored the possibility of achieving EIT in dressed qubit-resonator system due to the inherent tunability of the coherences.  Since the Autler-Townes effect is closely related to EIT \cite{Sillanpaa2009, Ian2010, Abdumalikov2010}, our observation of the Autler-Townes effect in a dressed qubit-resonator system is a step in the direction, though experimental realization of EIT in such a system still remains an open question.

 \ack
 {F. C. W. would like to acknowledge support from the Joint Quantum Institute and the State of Maryland through the Center for Nanophysics and Advanced Materials. The authors  would like to thank K. Osborn, M. Stoutimore, V. Zaretskey, R. Budoyo, B. K. Cooper, and J. B. Hertzberg for useful discussions.}

\section*{References}
\bibliography{atbibfinaljun2013}

\begin{thebibliography}{10}

\bibitem{Clarke2008}
Clarke John and Wilhelm~Frank K.
\newblock {\em Nature}, \textbf{453}(7198):1031--42, 2008.

\bibitem{Devoret2004a}
Devoret~M H and Martinis~J M.
\newblock {\em Quantum Information Processing}, \textbf{3}(October):163--203,
  2004.

\bibitem{Chang2013}
Chang J, Vissers~M R, Corcoles~A D, Sandberg M, Gao J, Abraham~D W, Chow~J M,
  Gambetta~J M, Rothwell~M B, Keefe~G A, Steffen M, and Pappas~D P.
\newblock {\em arXiv preprint arXiv:1303.4071}, pages 3--6, 2013.

\bibitem{Manucharyan2009}
Manucharyan~Vladimir E, Koch Jens, Glazman~Leonid I, and Devoret~Michel H.
\newblock {\em Science}, \textbf{326}(5949):113--6, 2009.

\bibitem{Paik2011}
Paik Hanhee, Schuster~D I, Bishop~Lev S, Kirchmair G, Catelani G, Sears~A P,
  Johnson~B R, Reagor~M J, Frunzio L, Glazman~L I, Girvin~S M, Devoret~M H, and
  Schoelkopf~R J.
\newblock {\em Phys. Rev. Lett.}, \textbf{107}(24):240501, 2011.

\bibitem{Barends2013}
Barends R, Kelly J, Megrant A, Sank D, Jeffrey E, Chen Y, Yin Y, Chiaro B,
  Mutus J, Neill C, O'Malley P, Roushan P, Wenner J, White~T C, Cleland~A N,
  and J~M Martinis.
\newblock {\em arXiv preprint arXiv:1304.2322}, (c):10, 2013.

\bibitem{Corcoles2011}
C{\'o}rcoles~Antonio D, Chow~Jerry M, Gambetta~Jay M, Rigetti Chad, Rozen~J R,
  Keefe~George A, {Beth Rothwell} Mary, Ketchen~Mark B, and Steffen M.
\newblock {\em App. Phys. Lett.}, \textbf{99}(18):181906, 2011.

\bibitem{Wenner2013}
Wenner J, Yin Yi, Lucero Erik, Barends R, Chen Yu, Chiaro B, Kelly J, Lenander
  M, Mariantoni Matteo, Megrant A, Neill C, O'Malley P~J J, Sank D, Vainsencher
  A, Wang H, White~T C, Cleland~A N, and Martinis~J M.
\newblock {\em Phys. Rev. Lett.}, \textbf{110}(15):150502, 2013.

\bibitem{Hood2000}
Hood~C J, Lynn~T W, Doherty~A C, Parkins~A S, and Kimble~H J.
\newblock {\em Science}, \textbf{287}(5457):1447--1453, 2000.

\bibitem{Raimond2001}
Raimond~J M, Brune M, and Haroche S.
\newblock {\em Rev. Mod. Phys.}, \textbf{73}(3):565--582, 2001.

\bibitem{Blais2004}
Blais Alexandre, Huang Ren-Shou, Wallraff Andreas, Girvin~S M, and R~J
  Schoelkopf.
\newblock {\em Phys. Rev. \textup{A}}, \textbf{69}(6):062320, 2004.

\bibitem{Wallraff2004}
Wallraff A, Schuster~D I, Blais A, Frunzio L, Huang R-S, Majer J, Kumar S,
  Girvin~S M, and Schoelkopf~R J.
\newblock {\em Nature}, \textbf{431}(7005):162--7, 2004.

\bibitem{Koch2007}
Koch Jens, Yu~Terri M, Gambetta Jay, Houck~A A, Schuster~D I, Majer J, Blais
  Alexandre, Devoret~M H, Girvin~S M, and Schoelkopf~R J.
\newblock {\em Phys. Rev. \textup{A}}, \textbf{76}(4):42319, 2007.

\bibitem{Schuster2007}
Schuster~D I, Houck~A A, Schreier~J A, Wallraff A, Gambetta~J M, Blais A,
  Frunzio L, Majer J, Johnson B, Devoret~M H, Girvin~S M, and Schoelkopf~R J.
\newblock {\em Nature}, \textbf{445}(7127):515--8, 2007.

\bibitem{You2011}
You~J Q and Nori Franco.
\newblock {\em Nature}, \textbf{474}(7353):589--97, 2011.

\bibitem{Dykman1987}
Dykman~M I and Krivoglaz~M A.
\newblock {\em Fiz. Tverd. Tela}, \textbf{29}(2):368--376, 1987.

\bibitem{Autler1955}
Autler~S H and Townes~C H.
\newblock {\em Phys. Rev.}, \textbf{100}(2):703--722, 1955.

\bibitem{Baur2008}
Baur M, Filipp S, Bianchetti R, Fink J, G\"{o}ppl M, Steffen L, Leek P, Blais
  A, and Wallraff A.
\newblock {\em Phys. Rev. Lett.}, \textbf{102}(24):243602, 2009.

\bibitem{Sillanpaa2009}
Sillanp\"{a}\"{a} Mika, Li~Jian, Cicak Katarina, Altomare Fabio, Jae Park,
  Raymond Simmonds, G~Paraoanu, and Pertti Hakonen.
\newblock {\em Phys. Rev. Lett.}, \textbf{103}(19):193601, 2009.

\bibitem{Bishop2010a}
Bishop~Lev S.
\newblock {\em {Circuit Quantum Electrodynamics}}.
\newblock PhD thesis, Yale University, 2010.

\bibitem{Jaynes1963}
Jaynes~E T and Cummings~F W.
\newblock {\em Proc. IEEE}, \textbf{51}(1):89--109, 1963.

\bibitem{Boissonneault2008a}
Boissonneault Maxime, Gambetta~J M, and Blais Alexandre.
\newblock {\em Phys. Rev. \textup{A}}, \textbf{79}(1):013819, 2009.

\bibitem{Gambetta2006}
Gambetta Jay, Blais Alexandre, Schuster~D I, Wallraff A, Frunzio L, Majer J,
  Devoret~M H, Girvin~S M, and Schoelkopf~R J.
\newblock {\em Phys. Rev. \textup{A}}, \textbf{74}(4):042318, 2006.

\bibitem{Boissonneault2010}
Boissonneault Maxime, Gambetta~J M, and Blais Alexandre.
\newblock {\em Phys. Rev. Lett.}, \textbf{105}(10):100504, 2010.

\bibitem{Carbonaro1979}
Carbonaro P, Compagno G, and Persico F.
\newblock {\em Phys. Lett. A}, \textbf{73}(2):97--99, 1979.

\bibitem{Glauber1963}
Glauber~Roy J.
\newblock {\em Phys. Rev.}, \textbf{131}(6):2766--2788, 1963.

\bibitem{Lindblad1976}
Lindblad G.
\newblock {\em Commun. math. Phys}, \textbf{48}:119--130, 1976.

\bibitem{Kim2011}
Kim Z, Suri B, Zaretskey V, Novikov S, Osborn~K D, Mizel A, Wellstood~F C, and
  Palmer~B S.
\newblock {\em Phys. Rev. Lett.}, \textbf{106}(12):120501, 2011.

\bibitem{Reed2010}
Reed~M D, DiCarlo L, Johnson~B R, Sun L, Schuster~D I, Frunzio L, and
  Schoelkopf~R J.
\newblock {\em Phys. Rev. Lett.}, \textbf{105}(17):173601, 2010.

\bibitem{Bishop2010}
Bishop~Lev S, Ginossar Eran, and Girvin~S M.
\newblock {\em Phys. Rev. Lett.}, \textbf{105}(10):100505, 2010.

\bibitem{Wallraff2007}
Wallraff A, Schuster~D I, Blais A, Gambetta~J M, Schreier J, Frunzio L,
  Devoret~M H, Girvin~S M, and Schoelkopf~R J.
\newblock {\em Phys. Rev. Lett.}, \textbf{99}(5):50501, 2007.

\bibitem{Boller1991}
Boller~K J, Imamo{\u g}lu A, and Harris~S E.
\newblock {\em Phys. Rev. Lett.}, \textbf{66}(20):2593--2596, 1991.

\bibitem{Ian2010}
Ian Hou, Liu Yu-xi, and Nori Franco.
\newblock {\em Phys. Rev. \textup{A}}, \textbf{81}(6):063823, 2010.

\bibitem{Hau1999}
Hau~L V, Harris~S E, Dutton Zachary, and Behroozi~Cyrus H.
\newblock {\em Nature}, \textbf{397}(February):594--598, 1999.

\bibitem{Murali2004}
Murali K V~R M, Dutton Z, Oliver~W D, Crankshaw~D S, and Orlando~T P.
\newblock {\em Phys. Rev. Lett.}, \textbf{93}(8):087003, 2004.

\bibitem{Abdumalikov2010}
Abdumalikov~A A, Astafiev O, Zagoskin~A M, Pashkin~Yu A, Nakamura Y, and Tsai~J
  S.
\newblock {\em Phys. Rev. Lett}, \textbf{104}(19):193601, 2010.

\end{thebibliography}
\bibliographystyle{unsrt}


\end{document}